# Multifunctional Lightweight Radiators for Small-Satellite Thermal Control

Karl Pederson[1], Sam Keller[1], Daniel Kindem[1], Hayden Hommes[1], Ognjen Ilic[1*]

[1] Department of Mechanical Engineering, University of Minnesota, Minneapolis, 55455 USA

**ABSTRACT**

Thermal management in small satellites is constrained by limited radiative area and strict mass budgets, necessitating the development of radiator structures that are simultaneously lightweight, thermally conductive, and mechanically robust. Here, we present a topology-optimization and design-space analysis framework for multifunctional lightweight radiators that achieve high specific stiffness and high effective thermal conductivity through simultaneous structural and thermal optimization. Density-based optimization produces hierarchical architectures that naturally form continuous cavities suitable for high-conductivity channels such as embedded heat pipes. The resulting microarchitectures exhibit Pareto behavior indicating efficient trade-offs between mass, stiffness, and thermal conductivity, while maintaining dynamic stability across a broad range of design parameters. Coupled structural-thermal analysis shows that voids used as thermal channels yield nearly isothermal radiating surfaces, confirming efficient lateral and transverse heat flow through the radiator. This integrated framework contributes toward the development of thermo-mechanically optimized radiator panels for small-scale spacecraft, enabling compact and efficient thermal control solutions.

Lightweight structures that combine high thermal conductivity and mechanical stiffness are increasingly important for operation in extreme environments, particularly for spacecraft and satellite applications. In orbit, spacecraft can experience drastic and repeated changes in the radiative environment, transitioning between intense solar irradiation and eclipse and producing large thermal fluctuations that must be managed with limited radiative surface area. For small satellites and microsystems such as CubeSats [1], these constraints are even more severe: small system size limits the available area for heat rejection, while ever-increasing onboard power requirements necessitate efficient heat conduction from the inside of the spacecraft bus to the outside space [2,3]. Deployable radiator panels—thin, large-area structures which are usually body-mounted or extendable—are widely used to increase radiating area and enhance thermal dissipation [4–6]. However, these radiators must simultaneously satisfy demanding mechanical and thermal criteria: they must maintain structural rigidity while minimizing mass, provide high thermal conductivity, and withstand dynamic vibrations during launch and in orbit. Traditional designs often employ sandwich composites with lightweight honeycomb cores to reduce density and maintain stiffness [7]. However, while such panels conduct heat reasonably well through the thickness, they have limited lateral heat spreading capabilities, so additional components are typically required to achieve sufficient thermal conductivity [8,9]. To improve heat transport and temperature uniformity, heat pipes and vapor chambers can be integrated into radiator panels, producing radiators that can closely approach isothermal behavior under operating conditions while maintaining high structural efficiency [8,10–12]. Yet, such hybrid approaches typically rely on the additive integration of mechanical and thermal subsystems rather than their co-design. In many spacecraft radiator designs, approaches have focused on parameter sweeps within predefined, parameterized design spaces, such as truss or lattice cores [13], without accessing the broader topology-optimization design space to address the tight size and weight constraints inherent to small-satellite and CubeSat platforms.

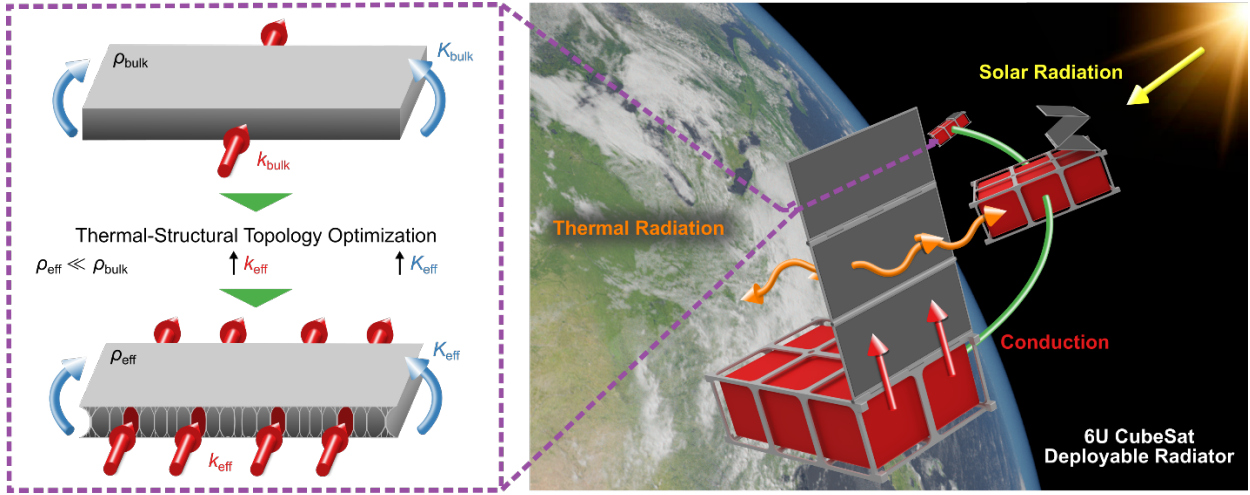

*Figure 1.* Conceptual illustration of a multifunctional radiator design: topology-optimized microarchitectures can achieve high effective stiffness ($K_{eff}$) and high thermal conductivity ($k_{eff}$) at a fraction of the density of a solid material ($\rho_{eff}$), enabling lightweight deployable radiators for small spacecraft, such as the illustrated 6U CubeSat.

In this work, we introduce a combined topology-optimization and design-space analysis framework for multifunctional radiator architectures tailored to small-satellites and CubeSat platforms (**Figure 1**). In a systematic analysis, this approach reveals lightweight structures with high specific stiffness with naturally forming continuous cavities that can host embedded heat pipes, thereby enabling high effective thermal conductivity at low density. We implement this approach using density-based topology optimization [14–16] to design microarchitected radiator profiles that maximize specific stiffness (stiffness normalized by density) while forming hierarchical "tree-like" motifs that concentrate material along principal load paths and enhance both mechanical stiffness and thermal conduction pathways.

Our manuscript is organized as follows. First, we systematically explore the relationship between radiator thickness, volume fraction, and material type, revealing mass-efficiency trends and mass/stiffness/thermal conductivity Pareto fronts that guide lightweight design for CubeSat-compatible radiators. Next, we evaluate the optimized architectures for bending stiffness, vibration response, and heat-transfer performance. Third, through coupled structural–thermal analysis, we show that voids formed during optimization can act as natural conduits for embedded heat pipes, leading to nearly isothermal radiator surfaces that radiate >99% of the power of an ideal blackbody at the same temperature. Finally, we compare free-geometry and constrained-geometry optimizations (where fixed circular cavities represent standardized heat-pipe profiles) to demonstrate that superior thermal uniformity can be achieved without compromising dynamic stability.

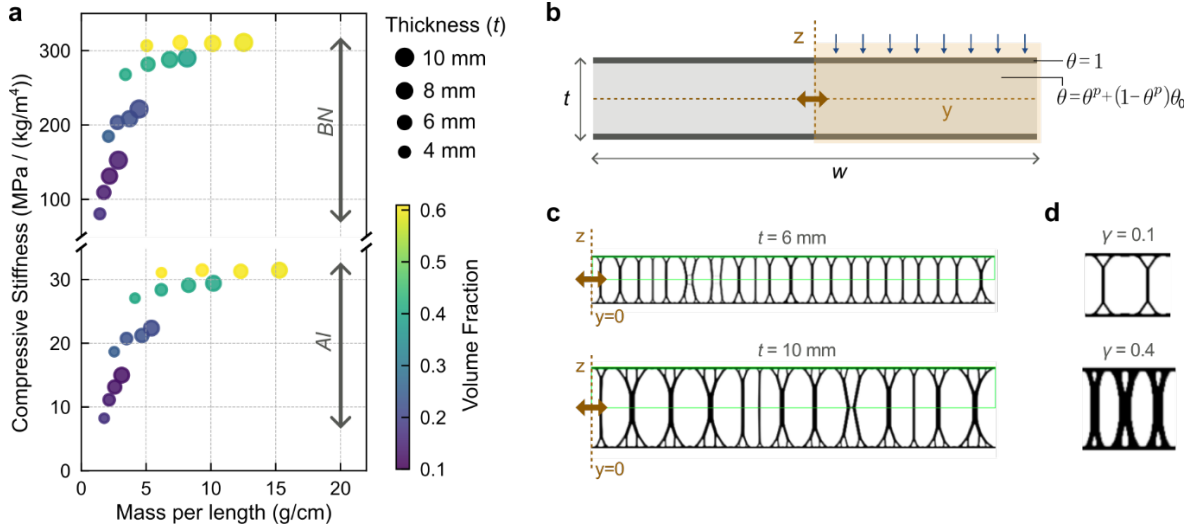

*Figure 2. (a) Pareto front of specific compressive stiffness versus mass-per-length for two representative material systems: a high-modulus ceramic (boron nitride, upper axis) and a moderate-modulus but thermally conductive metal alloy (Al-Li 8090, lower axis), illustrating the influence of material class on stiffness–mass trade-offs. Marker size represents radiator thickness, and color indicates solid volume fraction. (b) The radiator cross-section serves as a topology-optimization domain with fixed-density top and bottom layers and a SIMP-optimized central region. Here, θ denotes the density variable. (c) Representative optimized half-geometries for 6 mm and 10 mm thick radiator (mirror plane at y = 0), highlighting the emergence of "tree"-like motifs. (d) Enlarged views of topology patterns for volume-fraction constraints of 0.10 (top) and 0.40 (bottom), illustrating the influence of porosity on feature morphology.*

## RESULTS

We begin by using two-dimensional topology optimization to design radiator geometries that achieve high specific stiffness, defined as stiffness normalized by the effective density of the structure. **Figure 2** illustrates the approach taken to design two-dimensional geometries. In this formulation, a cross-section (y-z plane) is optimized to minimize the functional $J(\theta) = \int_\Omega \boldsymbol{\sigma} : \boldsymbol{\varepsilon} d\Omega$, which corresponds to minimizing structural compliance and therefore maximizing stiffness under compressive loading. Here, compressive stiffness serves as a proxy metric for bending stiffness, because it captures the material distribution's ability to resist deformation while remaining computationally tractable. We employ the solid isotropic material with penalization (SIMP) method of topology optimization implemented in COMSOL Multiphysics [17] (see Methods section). For a given unit cell, symmetry is enforced across both horizontal and vertical axes. The top and bottom faces of the unit cell are fixed to be solid (density θ = 1), while the central region has density $\theta = \theta^p + (1 - \theta^p)\theta_0$ from SIMP [14,18]. Specific stiffness is obtained from the equivalent homogenized modulus of the optimized cross-section divided by the effective density (determined from the spatially averaged material density profile) $\left(\frac{E_{eff}}{\rho}\right) = \frac{E_{eff} A_\Omega}{\int_\Omega \rho dA}$. For a fixed radiator width $w = 10$ cm, radiator thickness $t = $ (4 mm, 6 mm, 8 mm, 10 mm), volume fraction constraint $\gamma = $ (0.1, 0.2, 0.4, 0.6), and material (boron nitride [19], Al-Li 8090 alloy [20]) are varied. These parameter ranges are chosen to represent the practical design space for lightweight, thin radiators suitable for small satellite thermal management. As a result of topology optimization, tree-like motifs emerge. These motifs reflect the optimizer's tendency to concentrate material along principal load paths, forming hierarchical, branching structures. When thickness increases, fewer "tree" motifs are produced, leading to fewer but larger cavities (**Figure 2c**). Higher volume fractions lead to smaller cavities and less branching of small features (**Figure 2d**). However, as both thickness and volume fraction vary, these tree motifs are consistently present in the final structures. Because the resulting cavities are continuous, convex, and regularly spaced, they can serve as natural channels for embedding heat pipes or other high-conductivity pathways. Therefore, this optimization approach leads to structures with high specific stiffness and the capability to incorporate heat pipes to achieve high thermal conductivity.

We next evaluate the performance of the optimized geometries by examining the relationship between specific compressive stiffness and mass, revealing key tradeoffs across design parameters. This relationship, shown in **Figure 2a** as a Pareto front, highlights how design variables influence specific stiffness and mass efficiency. At low volume fractions, increasing radiator thickness leads to significantly higher specific stiffness with little mass penalty. In contrast, at high volume fractions, radiator thickness has little impact on stiffness while incurring significant added mass. A similar trend appears when changing the volume fraction at different thicknesses: increasing infill at low thickness has a significant positive impact with a small increase in mass; however, the mass penalty becomes exacerbated when increasing volume fraction at high thicknesses. These results indicate that mass efficiency decreases rapidly once the structure becomes relatively dense, as additional material no longer contributes meaningfully to load-bearing capacity. We note that changing the structural properties of the material has little impact on the optimized topologies or overall trends, suggesting that geometric design, rather than intrinsic material modulus, dominates the stiffness trends in this regime. The results suggest volume fraction is the dominant factor in determining the specific stiffness of the material. From a design standpoint, these findings imply that low-to-intermediate radiator thicknesses combined with moderate volume fractions achieve an optimal balance between stiffness and mass, offering a practical guideline for lightweight radiator design.

To investigate how the optimized geometries perform under bending, we extend our analysis to three-dimensional extrusions of the radiator cross-sections. **Figure 3a** visualizes the parameters of interest, which now include varying length ($L$), alongside thickness ($t$), and volume fraction ($\gamma$). This three-dimensional extrusion allows us to capture deformation and stiffness characteristics that cannot be fully represented in a two-dimensional compression analysis. **Figure 3b** shows the relationship between specific bending stiffness and mass for topology-optimized cross-sections extruded to various lengths. Here, low volume fractions result in both higher specific stiffness and less mass. This behavior indicates that mass-efficient designs naturally emerge at low infill levels, where material is distributed primarily along load paths. This same trend is present with radiator thickness, effectively removing the tradeoff between stiffness and mass which was present under compression. When two-dimensional topologies optimized for compressive stiffness are subjected to bending boundary conditions, the solid, thick material becomes the lower performance limit (in contrast to upper limit for uniform compression). This inversion reflects the geometric inefficiency of solid sections in bending, where stiffness more strongly depends on the distribution of material about the neutral axis. The removal of the tradeoff between mass and specific stiffness renders fabrication constraints and absolute bending stiffness requirements as the limiting factors. Among these, volume fraction (a common constraint in additive manufacturing) plays a particularly important role in determining achievable stiffness. **Figure 3c** shows a characteristic relationship between specific bending stiffness and volume fraction (here shown for a $t = 4$ mm thick, $L = 5$ cm long, radiator made of BN). Increasing the volume fraction from 0.1 to 0.6 results in approximately a 40% decrease in specific stiffness. This trend highlights the penalty of excessive infill, where added material increases mass without proportionally improving load-bearing efficiency. Low-thickness, low-volume-fraction radiator cross-sections provide high specific bending stiffness and low mass, enabling longer radiators for high-power heat rejection.

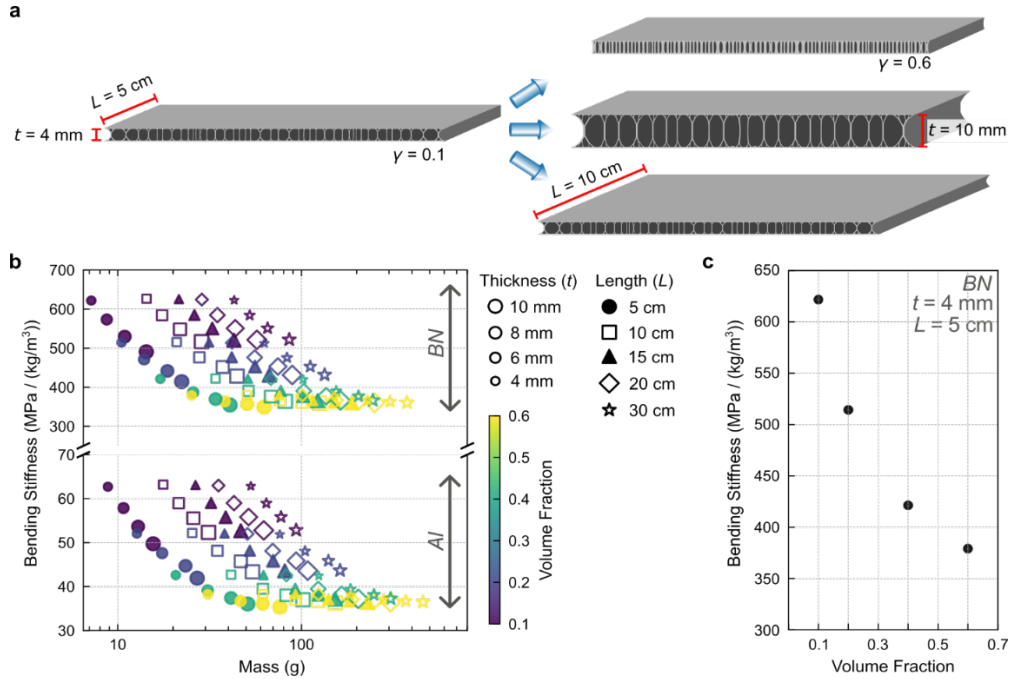

*Figure 3. Analysis of an extruded radiator. (a) Parameter design space includes radiator thickness (t), extruded radiator length (L), and volume fill-fraction (γ). (b) Specific bending stiffness versus mass for extruded radiator cores composed of a high-modulus ceramic (boron nitride, upper axis) and a moderate-modulus but thermally conductive metal alloy (Al-Li 8090, lower axis). Marker size corresponds to radiator thickness, color denotes volume fill-fraction, and marker shape indicates extrusion length. The visualized trend indicates that stiff and mass-efficient designs naturally emerge at low infill levels. (c) Relationship between specific bending stiffness and volume fraction (here shown for a t = 4 mm thick, L = 5 cm long, BN radiator). Increasing the volume fraction from 0.1 to 0.6 results in approximately a 40% decrease in specific stiffness.*

Next, we analyze the vibration response of optimized radiator structures to assess their dynamic stability and potential for resonance-related failure modes. **Figure 4a** depicts the profile of the three lowest vibrational modes of a microarchitected radiator with a clamped edge. While stiffness is an important design metric for structural integrity, natural frequency considerations are also important, as low resonant frequencies can amplify dynamic loads or induce fatigue, particularly in long, lightweight radiators subjected to spacecraft vibrations. **Figure 4b** shows the Pareto front between the fundamental (lowest mode) eigenfrequency and mass for various radiator thickness, volume fraction, and radiator lengths. Of these three parameters, we observe that the infill of a cross-section has the lowest impact on the fundamental mode frequency. Raising the volume fraction causes a small decrease in eigenfrequency, and thus is undesirable when designing for vibrations. Radiator thickness has a positive correlation with eigenfrequency, but the benefit is less pronounced with an increase in mass, highlighting the challenge of maintaining both low mass and high natural frequency. The most influential factor on the fundamental frequency of an extended radiator is its length. **Figure 4c** highlights this impact for an example of a 4 mm thick, 10% infill radiator. Here, the fundamental frequency of a long (30 cm) radiator is almost 100x lower than that of a short (5 cm) radiator. We note that the fundamental frequency of a long radiator can be increased by increasing its thickness. However, this improvement in frequency can lead to an additional penalty in mass efficiency. Therefore, optimizing for vibration resistance requires a balance between structural geometry and mass efficiency, with length emerging as the dominant design constraint.

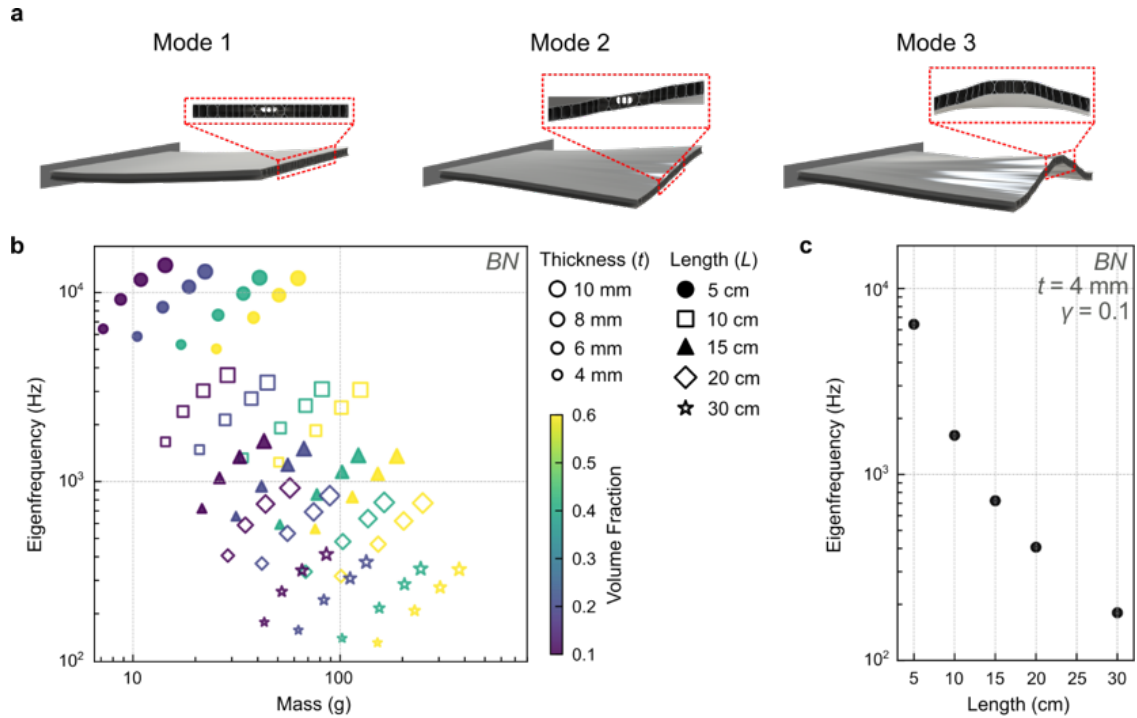

*Figure 4. Dynamic response of topology-optimized radiators. (a) Representative vibrational mode shapes of a radiator for the first three lowest frequency modes. Dynamic response considerations are important, as low resonant frequencies can amplify dynamic loads or induce fatigue, particularly in long, lightweight radiators subjected to spacecraft vibrations. (b) Pareto front of the fundamental (1st mode) eigenfrequency versus mass for a radiator of varying structural parameters. Marker size represents radiator thickness, color denotes solid-volume fraction, and shape indicates radiator length. (c) Fundamental eigenfrequency as a function of radiator length for an example radiator design, highlighting the strong length dependence.*

We examine the heat transfer potential of the optimized radiator structures with a heat transfer analysis in which voids between tree-like motifs are treated as embedded heat pipes, as illustrated in **Figure 5**. The selected voids are spaced to maintain an approximately uniform separation along the width of the radiator (details in Methods). The analysis models the upper-right quadrant of the structure, with symmetry assumed across both the horizontal and vertical axes. As the number of heat pipes $N_{hp}$ in the radiator increases, the temperature distribution along the radiator face becomes more uniform (**Figure 5a**). This behavior reflects improved lateral heat spreading as additional heat pipes provide efficient conduction pathways through the structure. **Figure 5b** compares the radiated power from the top surface to that of an isothermal surface, as a function of the number of voids ($N_{hp}$) selected as heat pipes. When the number of embedded heat pipes is low, conduction through the microarchitected lattice strongly influences the amount of radiated power and the limited conduction pathways lead to localized temperature gradients. As the number of heat pipes increases, conduction through them begins to dominate the system. Beyond approximately three embedded heat pipes, the radiator approaches a nearly isothermal condition, marking a transition from lattice-dominated to heat-pipe-dominated thermal transport. This shift results in a 25–30% increase in emitted power as the number of heat pipes increases from one to three. At higher heat pipe counts, the surface becomes effectively isothermal, radiating more than 99% of the power of an isothermal blackbody surface at $N_{hp} = 11$. Together, these results demonstrate that selectively utilizing cavities formed during topology optimization enables multifunctional structures that combine high mechanical stiffness with nearly uniform thermal performance.

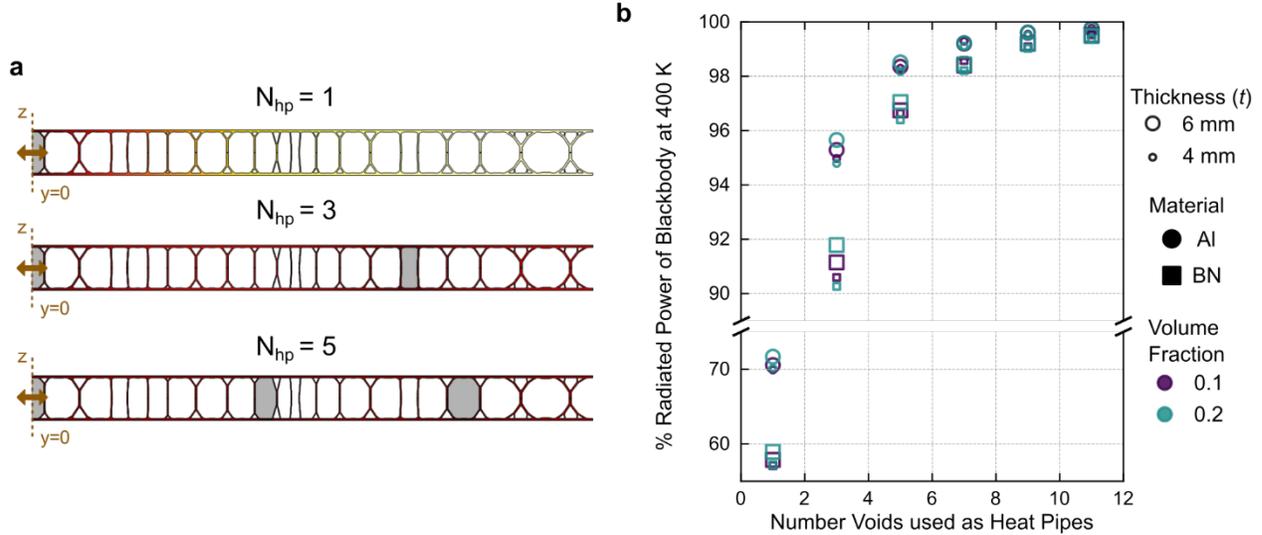

*Figure 5. Thermal performance of topology-optimized radiators with embedded heat channels. (a) Temperature profiles for the right half of the radiator as the number of heat channels increases (shaded regions denote voids treated as isothermal heat channels/pipes). Dark red indicates higher temperature and white indicates lower temperature, showing higher temperature uniformity with additional heat channels. The left half radiator temperature profile (not shown) is mirror-symmetric. (b) Radiated power from the surface as a function of the number of embedded heat channels. The power is normalized to the radiated power of an isothermal blackbody surface at the same temperature (400 K). Marker size represents radiator thickness, shape indicates base material, and color denotes volume fraction. Visualized trends show that selectively utilizing cavities formed during topology optimization as heat channels could enable very high effective thermal conductance.*

To compare the previous free-geometry optimization with a more constrained design strategy, we perform a topology optimization that incorporates fixed, circular heat pipe cavities at prescribed locations. We adopt this constrained optimization approach because, although the convex cavities in **Figure 5** could in principle function as heat channels, their varying and irregular shapes and sizes could complicate fabrication standardization. In an alternative approach, **Figure 6a** shows a topology optimization that constrains the geometry by fixing circular cavities at specific positions to serve as thermal channels/heat pipes (see Methods). Here, rather than selecting arbitrary cavities between tree motifs as thermal channels, conventional circular cavities are defined in advance. This modification enables direct control over the spacing and orientation of thermal channels while maintaining a structurally optimized surrounding architecture. Outside the protected circular cavity regions where the density variable is fixed, the microarchitecture is optimized in two dimensions for specific stiffness. **Figure 6b** shows the results of the constrained optimization. We observe that as the number of heat pipe cavities increases, the specific compressive stiffness of the structure decreases. This reduction in compressive stiffness is expected because, in essence, prescribing circular voids removes material that would otherwise contribute to compressive load paths. As the number of circular cavities increases, we also observe a substantial increase in radiated power. Similar to the topologies shown in **Figure 5**, a greater number of circular cavities acting as heat pipes produces an approximately isothermal top surface that achieves >99% of a blackbody surface's emission at the same temperature. However, in this case, the near-isothermal surface condition is achieved more efficiently because of the regular spacing and consistent geometry of the fixed cavity regions.

Next, we extrude the topologies with fixed circular cavities along the radiator length, and we analyze the bending stiffness and the dynamic response. Similar to the trend observed in **Figure 3**, the specific bending stiffness of radiators with circular voids inversely correlates with thickness and volume fraction (**Figure 6c**). As the number of circular voids increases, the specific bending stiffness increases slightly, suggesting an interplay between mass distribution and bending stiffness in radiators with circular cavities. This effect arises because the circular cavities act as stiffening features that preserve load-bearing pathways in bending, countering the stiffness reduction seen in compression. **Figure 6d** shows the general trends of the fundamental frequency for radiators with varying lengths, thicknesses, volume fractions, and number of circular cavities. As before, the fundamental frequency correlates positively with radiator thickness (i.e., thicker radiators have a higher fundamental mode resonant frequency) and negatively with volume fraction and radiator length. Interestingly, the fundamental frequency does not appear to be very sensitive to the number of circular cavities, suggesting that the dynamic response is primarily governed by global geometry and effective properties rather than localized cavity arrangement. These results collectively indicate that by prescribing protected circular cavity regions during optimization, it is possible to achieve superior thermal uniformity without compromising dynamic stability.

In summary, we have introduced a broad thermomechanical design-space analysis that leverages topology optimization to develop multifunctional lightweight radiators that address the competing requirements of stiffness, mass efficiency, and thermal conductivity in small-satellite platforms. Our approach combined coupled structural and thermal optimization to generate hierarchical microarchitectures that naturally form continuous cavities suitable for high-conductivity channels. Parametric investigation revealed clear Pareto trade-offs among mass, stiffness, and thermal performance while capturing dynamic considerations associated with the vibrational modes of the radiator. We envision several relevant extensions of this work. While in this work we focused on two relevant material examples (i.e., a high conductivity and a high stiffness material) to highlight the dominant trends, it would be relevant to expand this analysis to include advanced metal alloys and high-performance composites (for example, such as those integrating high-conductivity and high-stiffness boron nitride nanotubes [21–23]). In addition, a fully coupled thermal-fluid analysis of embedded heat channels/pipes, incorporating the effects of working-fluid phase change, would more accurately capture coupled thermofluidic behavior and its influence on system performance. Another promising direction lies in the integration and optimization of radiator architectures within multi-segment deployable mechanisms, enabling compact stowage in small-satellite configurations. Finally, system-level integration studies that link radiator performance to spacecraft power, thermal, and structural subsystems would be critical for evaluating these multifunctional designs for flight-ready thermal control solutions.

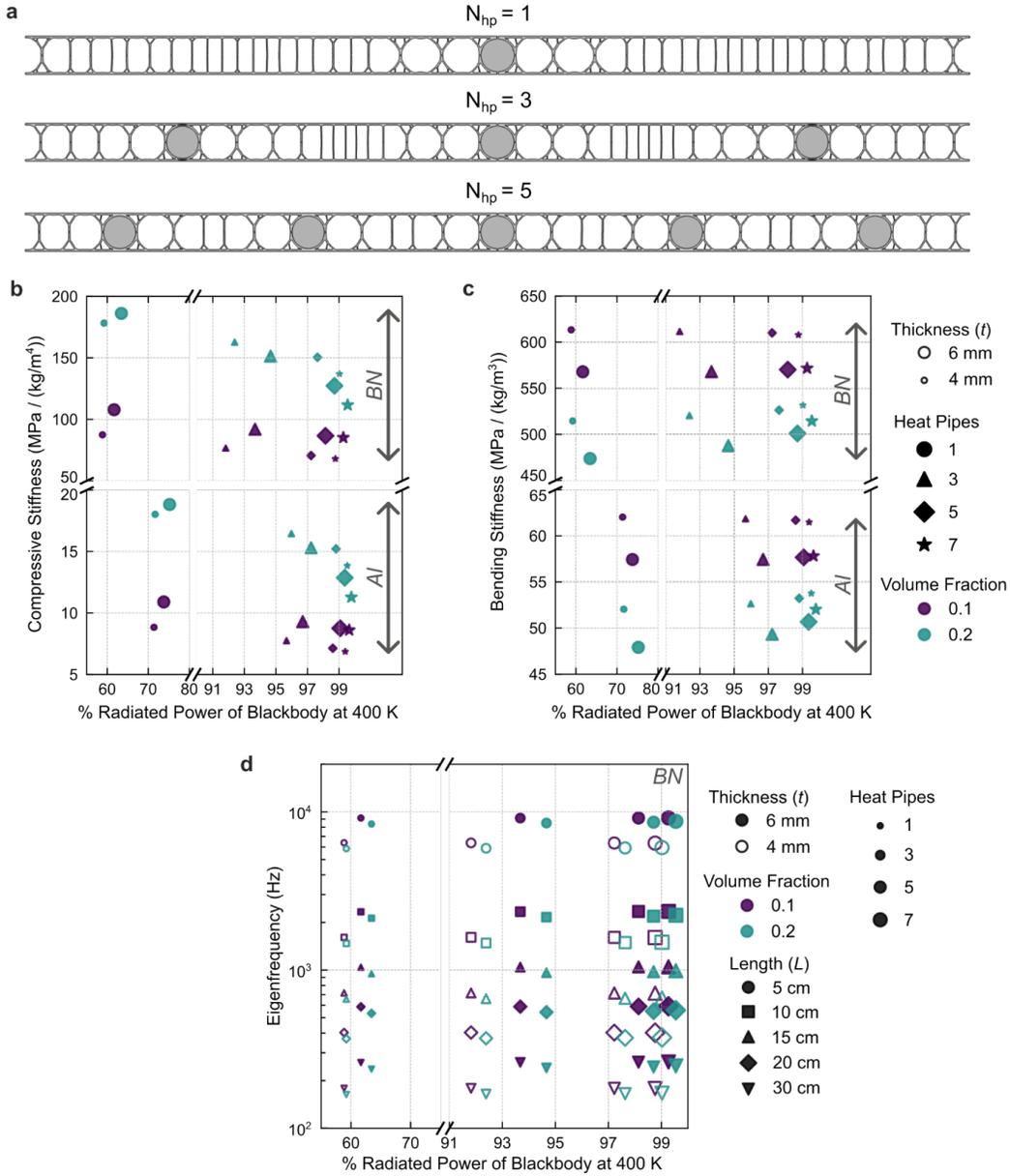

*Figure 6. Thermo-mechanical analysis of radiators obtained from constrained topology optimization with fixed circular heat-pipe cavities. (a) Optimized topologies for $N_{hp} = (1, 3, 5, 7)$ heat pipe channels, where each heat pipe channel is defined as a protected circular region excluded from the optimization domain. (b) Specific compressive stiffness versus normalized radiated power (ratio of emitted power to that of an isothermal blackbody), illustrating the trade-off between structural and thermal performance. (c) Specific bending stiffness versus normalized radiated power. Marker size represents radiator thickness, shape denotes the number of heat-pipe cavities, and color corresponds to volume fraction. (d) Fundamental eigenfrequency versus normalized radiated power. Here, marker fill encodes radiator thickness while shape indicates radiator length. These results show that prescribing circular heat-pipe regions enhances thermal uniformity with minimal compromise in stiffness or dynamic response.*

**Acknowledgements:** The authors thank R. Sweat for the helpful discussion. This material is based upon research supported by DARPA under Award Number D24AP00310-00.

*Email: ilic@umn.edu